\documentstyle[pre,aps,epsf,multicol,tabularx]{revtex}

\tighten
\begin{document}

\newcommand{\bfGamma}{\mbox{\boldmath $\bf\Gamma$}}
\newcommand{\bfq}{\mbox{\boldmath $\bf q$}}
\newcommand{\bfp}{\mbox{\boldmath $\bf p$}}

\newcommand{\smallbfGamma}{\mbox{\boldmath $\scriptstyle\Gamma$}}
\newcommand{\caGamma}{\mbox{$\it\Gamma$}}
\newcommand{\bfXi}{\mbox{\boldmath $\bf \Xi$}}

\baselineskip 0.382cm


\draft

\title{Master equation approach to the conjugate pairing rule 
   of Lyapunov spectra for many-particle thermostatted systems}

\author{Tooru Taniguchi and Gary P. Morriss}

\address{School of Physics, University of New South Wales, 
Sydney, New South Wales 2052, Australia}

\date{\today}

\maketitle

\vspace{-0.1cm}
\begin{abstract}

\baselineskip 0.37cm

   The master equation approach to Lyapunov spectra for 
many-particle systems is applied to non-equilibrium 
thermostatted systems to discuss the conjugate pairing rule. 
   We consider iso-kinetic thermostatted systems 
with a shear flow sustained by an external restriction,  
in which particle interactions are expressed as  
a Gaussian white randomness. 
   Positive Lyapunov exponents are calculated by using 
the Fokker-Planck equation to describe 
the tangent vector dynamics. 
   We introduce another Fokker-Planck equation to describe 
the time-reversed tangent vector dynamics, which allows us
to calculate the negative Lyapunov exponents. 
   Using the Lyapunov exponents provided by these two 
Fokker-Planck equations we show the conjugate 
pairing rule is satisfied for thermostatted 
systems with a shear flow in the thermodynamic limit. 
   We also give an explicit form to connect the Lyapunov 
exponents with the time-correlation of the interaction matrix 
in a thermostatted system with a color field. 

\end{abstract}

%

\vspace{0cm}

\vspace{0.3cm}

\begin{multicols}{2}

\narrowtext


%
%
%
%
%
%
%
%


\section{Introduction}

   The Lyapunov exponent is an essential concept to express the 
instability of orbits in a dynamical system. 
   It is introduced as an exponential expansion (or contraction) 
rate of an infinitesimal perturbation of orbits, and its positivity 
implies that the system is chaotic. 
   In general there is a Lyapunov exponent for each independent
direction of the infinitesimal perturbation of the orbit, 
and the sorted set of such Lyapunov exponents is called 
the Lyapunov spectrum, and has been the subject of study 
in many-particle systems. 
   For example, the existence of its thermodynamic limit 
\cite{Liv87,Sin96,Tan02}, an effect of the rotational 
degrees of freedom of molecules \cite{Mil98b}, its stepwise 
structure and the Lyapunov modes \cite{Pos00}, and 
a tracer particle effect \cite{Gas01} have been observed and 
discussed in the Lyapunov spectra of many-particle chaotic systems. 

   A characteristic of the Lyapunov spectrum that 
is known in Hamiltonian systems, is that the Lyapunov 
exponents appear as a pair, namely any positive Lyapunov exponent 
accompanies a negative Lyapunov exponent with its opposite sign 
\cite{Arn89}. 
   This characteristic, which is based on the symplectic 
structure of the Hamiltonian mechanics, is not correct 
in non-Hamiltonian 
systems, but it is interesting to know how it is modified 
in quasi-Hamiltonian systems such as a Hamiltonian system coupled 
to a thermodynamic reservoir. 
   This problem has been considered in some thermostatted dynamics 
with a term to extract the heat produced in the system 
by external force fields, and led to the proposal of the  
{\em conjugate pairing rule} for thermostatted systems, which claims 
that the sum of any Lyapunov exponent pair is not zero but a constant 
regardless of the exponent numbers \cite{Dre88,Eva90}. 
   This conjecture was confirmed by many following numerical 
calculations \cite{Sar92,Det95,Del96}.
   This also led to the discovery that some thermostatted systems 
contain hidden Hamiltonian structure \cite{Det96c,Woj98}. 
   The pairing rule is not only interesting as a mathematical 
structure of the thermostatted system 
but also valuable for a practical use; the conjugate 
pairing rule for the thermostatted system allows us to calculate 
non-equilibrium transport 
coefficients (eg. conductivity and viscosity) from 
only one pair of the Lyapunov exponents, such as the largest and 
smallest Lyapunov exponents only 
\cite{Eva90,Del96,Mor88,Bei96}.

   A problem is that the necessary and the sufficient conditions 
for the conjugate pairing rule to hold for thermostatted systems 
is not clearly known. 
   The conjugate pairing rule for the iso-kinetic thermostatted  
system with a color field was proved for the soft core interaction 
potential \cite{Det96a} and the hard core interaction potential 
\cite{Woj98,Pan01}, regardless of the number of particles. 
   A similar discussion was done in Nos{\'e} Hamiltonian dynamics 
\cite{Det97}. 
   These works give the sufficient conditions for the conjugate 
pairing rule. 
   On the other hand, it was suggested numerically that 
it can be violated in the presence 
of a magnetic field \cite{Dol00} and in inhomogenously thermostatted 
systems such as a system under temperature gradient 
\cite{Wag00} or a system in which the peculiar momenta are  
thermostatted \cite{Sea98}. 
   Another numerical work also suggested that it is not exact in 
the iso-energetic thermostat with a finite number of particles 
\cite{Bon98}, although the iso-energetic thermostat should be   
equivalent to the iso-kinetic thermostat as the number of 
the particles goes to infinity, namely in the thermodynamic limit 
\cite{Eva93,Rue00}.  
   A special interest is the iso-kinetic thermostatted system 
with a shear field, which is described by the Sllod equation 
for the planar Coutte flow \cite{Eva90a}.
   Early investigations supported the conjugate pairing rule 
for such a system \cite{Eva90,Sar92,Gul94}. 
    Refs. \cite{Sea98,Isb97} suggested 
a small deviation from the conjugate pairing rule. 
   An analytical consideration showed that the deviation from 
the conjugate pairing rule should be at most fourth order 
in the shear rate in the case of a small shear rate in 
the thermodynamic limit \cite{Pan02a}.
   However a recent numerical calculation 
with a more careful numbering of the Lyapunov exponents 
and with numerical error bars showed that 
within the numerical precision  
the conjugate pairing rule was satisfied \cite{Mor02}. 
   After all these trials, a justification of the conjugate 
pairing rule for the iso-kinetic thermostatted system with a shear 
field still remains as an open problem. 

   Analytical calculation of the full Lyapunov spectra 
for many-particle systems is still not an easy matter, 
and so far full Lyapunov spectra have been calculated mainly 
using numerical approaches. 
   On the other hand, recently an analytical approach to the full 
Lyapunov spectra was proposed for many-particle systems with 
a random interaction \cite{Tan01}. 
   This approach describes the tangent vector dynamics by 
a master equation, and allows the calculation of all individual 
positive Lyapunov exponents through the average of the magnitude 
of the tangent vector. 
   It was used to explain the stepwise structure of the 
Lyapunov spectrum for many-particle Hamiltonian systems. 
   However a generalization of this approach to non-equilibrium 
systems, especially to thermostatted systems, is not known yet. 

   The purpose of this paper is to generalize the master 
equation approach to the Lyapunov spectrum to non-equilibrium  
thermostatted systems, and to discuss the conjugate pairing 
rule of the Lyapunov spectrum for the iso-kinetic thermostatted 
system with a shear field. 
   Especially we propose a method to calculate negative 
Lyapunov exponents using a time-reversed master equation. 
   In this paper we concentrate on the case where the particles 
interact with a Gaussian white randomness. 
    In this case the master equation that describes the tangent 
vector dynamics is attributed to the Fokker-Planck equation. 
   We also restrict our consideration in the thermodynamic limit, 
in which the fluctuations of the friction coefficient 
can be neglected \cite{Gul94}. 
   Under these conditions we show that the conjugate pairing 
rule for the thermostatted system with a shear field 
given by the Sllod equation is satisfied.
   As a special case we also discuss briefly an explicit form 
to connect the Lyapunov exponents with the time-correlation of 
the interaction matrix in a thermostatted system 
without a shear field. 





\section{Isokinetic Thermostatted System with a Shear Field 
and its Tangent Vector Dynamics}

   We consider non-equilibrium systems with an iso-kinetic 
thermostat. 
   Our consideration includes the case where a shear flow 
is sustained by an external restriction, and for simplicity 
we consider a 
two-dimensional system consisting of $N$ particles with the same 
mass $m$. 
   We introduce $\bfq^{(j)}(t) 
\equiv(q^{(j)}_{x}(t),q^{(j)}_{y}(t))^{T}$ 
and $\bfp^{(j)}(t)\equiv(p^{(j)}_{x}(t),p^{(j)}_{y}(t))^{T}$ as 
the spatial coordinate vector and the momentum vector of 
the $j$-particle, respectively, at time $t$ 
with the transpose operation $T$. 
   (Note that all vectors in this paper are 
introduced as column vectors.)
   Equations for $\bfq^{(j)}(t)$ and $\bfp^{(j)}(t)$ are 
expressed as \cite{Eva90a}

\begin{eqnarray}
   \frac{d\bfq^{(j)}(t)}{dt} = \frac{1}{m} \bfp^{(j)}(t) 
   + \gamma \Xi_{2} \bfq^{(j)}(t)
\label{ThermEquat1}\end{eqnarray}
\begin{eqnarray}
   \frac{d\bfp^{(j)}(t)}{dt} = - \frac{\partial U(t)}
   {\partial\bfq^{(j)}(t)} 
   - \left[\gamma \Xi_{2} + \alpha(t) I_{2} \right] \bfp^{(j)}(t)
\label{ThermEquat2}\end{eqnarray}

\noindent where $U(t)$ is the potential energy as a function 
of $\bfq^{(j)}(t), j=1,2,\cdots,N$ and $t$ only, and we introduce 
$\Xi_{2k}$ is the $2k \times 2k$ matrix defined by 

\begin{eqnarray}
   \Xi_{2k} \equiv 
   \left(
   \begin{array}{cc}
      0_{k} & I_{k}     \\
      0_{k} & 0_{k}   
   \end{array}
   \right)
\end{eqnarray}

\noindent with the $k \times k$ identical matrix $I_{k}$ 
and the $ k\times k$ null matrix $0_{k}$. 
   Here $\gamma$ is the shear rate as an external parameter,  
namely a constant gradient of the $x$ component of the local 
velocity in the $y$ direction, and $\alpha(t)$ is defined by 

\begin{eqnarray}
   \alpha(t) \equiv - 
   \frac{\sum_{j=1}^{N}\bfp^{(j)}(t)^{T} \left(\frac{\partial U(t)}
   {\partial\bfq^{(j)}(t)} + \gamma \Xi_{2} \bfp^{(j)}(t)\right)}
   {\sum_{j=1}^{N} |\bfp^{(j)}(t)|^{2}}
\label{ShearRate}\end{eqnarray}

\noindent so that the total kinetic energy is constant in time: 
$d\bigl[\sum_{j=1}^{N}|\bfp^{(j)}(t)|^{2}/(2m)\bigr]/dt = 0$. 
   Eqs. (\ref{ThermEquat1}) and (\ref{ThermEquat2}) are called 
the Sllod equation for the planar Coutte 
flow with the isokinetic thermostat, 
and gives the model for the system driven 
by external fields and/or a shear rate with an attached heat 
reservoir which removes the energy generated inside the system 
and maintains the temperature of the system constantly in time. 
   As an example described by Eqs. (\ref{ThermEquat1}) and 
(\ref{ThermEquat2}),  other than the system with a shear field, 
we may mention the color field system in which the system consists 
of many particles with charges of different signs and is driven 
by an external electric field \cite{Del96,Pos88}.  


   In general, the quantity $\alpha(t)$, which is interpreted as 
the friction coefficient, depends on the coordinates and the 
momenta of the particles, so is variable in time. 
   However, it is known that the fluctuation of the quantity 
$\alpha(t)$ is small in a system consisting of many particles 
\cite{Gul94}.
   (For a justification of this point by the kinetic approach 
see Ref. \cite{Zon99}, which shows that the quantity $\alpha(t)$ 
fluctuates with the order of $1/\sqrt{N}$ around a fixed value.) 
   Based on this fact, in this paper we consider only the system 
which consists of enough particles so that the friction 
coefficient $\alpha(t)$ in Eq. (\ref{ThermEquat2}) can be replaced 
by a fixed constant $\bar{\alpha}$. 
 
   For a convenience we represent the $4N$-dimensional phase 
space vector $\bfGamma(t)$ as a vector $(
q_{x}^{(1)}(t), q_{x}^{(2)}(t), \cdots, q_{x}^{(N)}(t), 
$ $q_{y}^{(1)}(t), q_{y}^{(2)}(t), \cdots, q_{y}^{(N)}(t), 
$ $p_{x}^{(1)}(t), p_{x}^{(2)}(t), \cdots, p_{x}^{(N)}(t), 
$ $p_{y}^{(1)}(t), p_{y}^{(2)}(t), \cdots, p_{y}^{(N)}(t) )^{T}$. 
   Using this notation and the assumption explained in the previous 
paragraph we obtain the equation 

\begin{eqnarray}
   \frac{d\delta\bfGamma(t)}{dt} = {\cal L}(t) \delta\bfGamma(t)
\label{LineaDynam}\end{eqnarray}

\noindent for the tangent vector $\delta\bfGamma(t)$. 
   Here the matrix ${\cal L}(t)$ is given by 
   
\begin{eqnarray}
   {\cal L}(t) \equiv 
   \left( \begin{array}{cc}
     \Phi &  I_{2N}/m      \\
     R(t) &  \Psi  
   \end{array}\right)
\label{MatriCalL}\end{eqnarray}

\noindent with $2N \times 2N$ matrices $\Phi$, $\Psi$ and 
$R$ defined by 

\begin{eqnarray}
   \Phi \equiv \gamma \Xi_{2N} ,
\label{MatriPhi}\end{eqnarray}
\begin{eqnarray}
   \Psi \equiv - \gamma \Xi_{2N} - \bar{\alpha} I_{2N} ,
\label{MatriPsi}\end{eqnarray}
\begin{eqnarray}
   R(t) \equiv 
   - \frac{\partial^{2} U(t)}{\partial\bfq(t)\partial\bfq(t)}
\label{RandoR}\end{eqnarray}

\noindent where we introduced $\bfq(t)$ as a vector $(
q_{x}^{(1)}(t), q_{x}^{(2)}(t), \cdots, $ $q_{x}^{(N)}(t), 
$ $q_{y}^{(1)}(t), q_{y}^{(2)}(t), \cdots, q_{y}^{(N)}(t))^{T}$.


\section{Random Interactions and Master Equations for the Tangent 
Vector Dynamics} 

In this section we introduce a random interaction between 
the particles, 
and obtain the two kinds of master equations 
corresponding to the time-forward tangent vector 
dynamics and the time-reversed tangent vector dynamics by 
using Kramers-Moyal expansion technique.

\subsection{Fokker-Planck equation 
for the forward dynamics of the tangent vector}

   We consider the case that each particle interacts with the other 
particles randomly enough so that the matrix 
$R(t)\equiv(R_{jk}(t))$ can be regarded as a Gaussian white random 
matrix satisfying the conditions 

\begin{eqnarray}
   \langle R_{\mu_{1}\nu_{1}}(t_{1}) 
      R_{\mu_{2}\nu_{2}}(t_{2})
      \cdots
      R_{\mu_{2n-1}\nu_{2n-1}}(t_{2n-1})\rangle = 0 , 
\label{WhiteNoise1}\end{eqnarray}
\begin{eqnarray}   
   && \langle R_{\mu_{1}\nu_{1}}(t_{1}) 
       R_{\mu_{2}\nu_{2}}(t_{2}) 
      \cdots
      R_{\mu_{2n}\nu_{2n}}(t_{2n})
      \rangle \nonumber \\
   && \hspace{0.6cm} =
      \sum_{P_{d}} 
      D_{\mu_{j_{1}}\nu_{j_{1}}\mu_{j_{2}}\nu_{j_{2}}} 
      D_{\mu_{j_{3}}\nu_{j_{3}}\mu_{j_{4}}\nu_{j_{4}}} 
     \nonumber \\
	 && \hspace{1cm}  \cdots     
     D_{\mu_{j_{2n-1}}\nu_{j_{2n-1}}\mu_{j_{2n}}\nu_{j_{2n}}} 
      \nonumber \\
   && \hspace{1cm} \times  
      \delta(t_{j_{1}}-t_{j_{2}})
      \delta(t_{j_{3}}-t_{j_{4}})
      \cdots
      \delta(t_{j_{2n-1}}-t_{j_{2n}})
\label{WhiteNoise2}\end{eqnarray}

\noindent for any integer $n$ and a $4$-th rank constant tensor 
$D_{jkln}$, where we take the sum over only the 
permutation $P_{d}:$ $ (1,2,\cdots$ $,2n)$ $\rightarrow$ $ 
(j_{1}, j_{2}, \cdots,j_{2n})$, and the bracket $\langle\cdots\rangle$ 
means the ensemble average over random processes.

   Under the randomness conditions (\ref{WhiteNoise1}) and (\ref{WhiteNoise2}) 
the time-evolutional equation (\ref{LineaDynam}) is regarded as a stochastic equation 
of the Langevin type, and its corresponding master equation 
for the the probability 
density $\rho^{(+)}(\delta\bfGamma,t)$ for the tangent vector $\delta\Gamma$ at 
time $t$ is given by 

\begin{eqnarray}
   &&
   \frac{\partial }{\partial t} 
      \; \rho^{(+)} (\delta\bfGamma,t)
	  \nonumber \\
   && \hspace{0cm} 
	  = -\sum_{\mu=1}^{2N} \sum_{\nu=1}^{2N} 
	   \frac{\partial}{\partial \delta q_{\mu}}
	   \left(\Phi_{\mu\nu} \delta q_{\nu} 
	   +\frac{\delta_{\mu\nu}}{m} \delta p_{\nu}\right)
       \rho^{(+)}(\delta\bfGamma,t)
       \nonumber \\
   && 
   \hspace{0.5cm} 
	  - \sum_{\mu=1}^{2N}\sum_{\nu=1}^{2N}  \frac{\partial}
	  {\partial \delta p_{\mu}} \;  \Psi_{\mu\nu} \delta p_{\nu} 
      \; \rho^{(+)}(\delta\bfGamma,t)
      \nonumber \\
   &&
   \hspace{0.5cm} 
      + \sum_{\mu=1}^{2N}\sum_{\nu=1}^{2N} 
      \sum_{\mu'=1}^{2N}\sum_{\nu'=1}^{2N}
      \frac{1}{2} D_{\mu'\mu \nu'\nu} \delta q_{\mu} 
      \delta q_{\nu} 
      \nonumber \\
   &&
   \hspace{3cm} \times\frac{\partial^{2}}{\partial 
      \delta p_{\mu'} \partial \delta p_{\nu'}} 
	  \; \rho^{(+)}(\delta\bfGamma,t)
\label{Fokke}\end{eqnarray}
 
\noindent applying the Kramers-Moyal expansion technique to 
the dynamics (\ref{LineaDynam}). 
   Here $\delta q_{j}$ and 
$\delta p_{j}$ are 
the $j$-th components of the coordinate part $\delta\bfq$ and 
the momentum part $\delta\bfp$ in the tangent 
vector $\delta\Gamma=(\delta\bfq,\delta\bfp)^{T}$, respectively, 
and $\Phi_{\mu\nu}$ and $\Psi_{\mu\nu}$ are 
the matrix elements of the matrix $\Phi$ and $\Psi$ 
defined by Eqs. (\ref{MatriPhi})  and (\ref{MatriPsi}), 
respectively. 
   The derivation of Eq. (\ref{Fokke}) is given in Appendix 
\ref{Maste}. 
   Eq. (\ref{Fokke}) in the special case of $\Phi=0_{2N}$ 
and $\Psi=0_{2N}$ have already been used to discuss the stepwise 
structure of the Lyapunov spectrum for a many-particle Hamiltonian 
system \cite{Tan01}.


\subsection{Anti-Fokker-Planck equation for the time-reversed 
dynamics of the tangent vector}

   As shown in Ref.  \cite{Tan01}, in the case of 
$\gamma=0$ and $\bar{\alpha}=0$ the Fokker-Planck equation 
(\ref{Fokke}) provides the positive Lyapunov exponents 
as the time averaged exponential rate of the randomness average 
by the probability density $\rho^{(+)}(\bfGamma,t)$
in the time evolution of infinitesimal perturbations of the 
dynamical variables. 
   However this method does not provide directly the 
negative Lyapunov exponents, because in the stochastic system 
{\em the randomness average of} the distance between the 
infinitesimal nearby trajectories should not shrink in the infinite 
time limit. 
   This was not a problem in the Hamiltonian system 
discussed in Ref. \cite{Tan01}, because in the Hamiltonian system 
the absolute values of the negative 
Lyapunov exponents are the same with the positive Lyapunov exponents 
\cite{Arn89}. 
   However in the thermostatted system discussed in this paper 
such a simple relation of the negative and 
the positive Lyapunov exponents can not be expected any more. 
   In order to overcome this problem and to provide the negative 
Lyapunov exponents using the master equation approach directly, 
we use 
the fact that {\em the negative Lyapunov exponents can be regarded 
as the positive Lyapunov exponents for the time-reversed motion}. 
   This fact has been used in some works to calculate 
the negative Lyapunov 
exponents for chaotic systems \cite{Bei96,Hoo88,Bei98}. 

   In the iso-kinetic thermostatted system with a shear field 
the time-reversed motion  
is expressed by the "time-reversed mapping" 
\cite{Eva90a}: 
$t \rightarrow -t$, $\bfq \rightarrow \bfq$, 
$\bfp \rightarrow - \bfp$ and $\gamma \rightarrow -\gamma$. 
   The transformation $\gamma \rightarrow -\gamma$ 
is justified by the fact that the direction of the shear flow 
changes to the opposite direction in the time-reversed motion. 
   This justifies the time-reversal transformation 
$\bar{\alpha} \rightarrow - \bar{\alpha}$ for the 
friction coefficient by Eq. (\ref{ShearRate}). 
   The time-reversed mapping leads to the time-reversed 
transformations 

\begin{eqnarray}
   t \rightarrow -t
      \label{TimeRever1} \\
   \delta\bfq \rightarrow \delta\bfq
      \label{TimeRever2} \\
   \delta\bfp \rightarrow -\delta\bfp
      \label{TimeRever3} \\
   \Phi \rightarrow -\Phi
	  \label{TimeRever4} \\
   \Psi \rightarrow -\Psi 
      \label{TimeRever5}
\end{eqnarray}

\noindent for the tangent vector dynamics, noting the 
relations (\ref{MatriPhi}) and (\ref{MatriPsi}) 
to connect the matrices $\Phi$ and $\Psi$ with 
the shear rate $\gamma$ and the friction coefficient 
$\bar{\alpha}$. 
   It is important to note that the tensor $D_{jkln}$ and 
the tangent dynamical equation (\ref{LineaDynam}) themselves 
are invariant under the transformations 
(\ref{TimeRever1}-\ref{TimeRever5}).

   Now we introduce the Fokker-Planck equation for 
the probability density 
$\rho^{(-)}(\delta\bfGamma,t)$ for the time-reversed tangent 
vector at time $t$
as the transformed equation of the Fokker-Planck equation 
(\ref{Fokke}) by  
the transformations (\ref{TimeRever1}-\ref{TimeRever5}), namely 

\begin{eqnarray}
   && \hspace{0cm} 
   \frac{\partial }{\partial t} 
      \; \rho^{(-)} (\delta\bfGamma,t)
	  \nonumber \\
   && \hspace{0cm} 
	   =  -\sum_{\mu=1}^{2N} \sum_{\nu=1}^{2N} 
	   \frac{\partial}{\partial \delta q_{\mu}}
	   \left(\Phi_{\mu\nu} \delta q_{\nu} 
	   +\frac{\delta_{\mu\nu}}{m} \delta p_{\nu}\right)
       \rho^{(-)}(\delta\bfGamma,t)
       \nonumber \\
   && 
   \hspace{0.5cm} 
	  - \sum_{\mu=1}^{2N}\sum_{\nu=1}^{2N}  \frac{\partial}
	  {\partial \delta p_{\mu}} \;  \Psi_{\mu\nu} \delta p_{\nu} 
      \; \rho^{(-)}(\delta\bfGamma,t)
      \nonumber \\
   &&
     \hspace{0.5cm} 
      - \sum_{\mu=1}^{2N}\sum_{\nu=1}^{2N} 
      \sum_{\mu'=1}^{2N}\sum_{\nu'=1}^{2N}
      \frac{1}{2} D_{\mu'\mu \nu'\nu} \delta q_{\mu} 
      \delta q_{\nu} 
      \nonumber \\
   &&
   \hspace{3cm} \times \frac{\partial^{2}}{\partial 
      \delta p_{\mu'} \partial \delta p_{\nu'}} \; 
	  \rho^{(-)}(\delta\bfGamma,t) .
\label{AntiFokke}\end{eqnarray}
  
\noindent This is simply the equation with the opposite sign 
of the diffusion term to the forward 
Fokker-Planck equation (\ref{Fokke}), 
and is interpreted as the master equation 
to describe the time-evolution of the tangent vector 
whose initial condition is the time-reversed initial condition 
to the Fokker-Planck equation (\ref{Fokke}).  
   We call this equation for the probability density 
$\rho^{(-)}(\delta\bfGamma,t)$ 
the "anti-Fokker-Planck equation" in this paper, and 
calculate the negative 
Lyapunov exponents as the exponential rate of the randomness 
average by the probability density 
$\rho^{(-)}(\delta\bfGamma,t)$
in the time evolution of infinitesimal perturbations 
of the dynamical variables in the minus infinite time. 

   It is important to note the difference between the 
anti-Fokker-Planck 
equation (\ref{AntiFokke}) and 
the so called "backward Fokker-Planck equation" \cite{Ris89}.
   The backward Fokker-Planck equation describes the dynamics 
before an initial time in which the initial condition is the same 
as in the forward Fokker-Planck equation 
to describe the dynamics after the initial time. 
   However, in order to calculate the negative Lyapunov exponents 
from the time-reversed motion, we must use the different initial 
condition which has the opposite sign of the momentum 
to the initial condition in the forward Fokker-Planck equation. 
   In the equilibrium case expressed by $\gamma=0$ and 
$\bar{\alpha}=0$ 
the backward Fokker-Planck equation coincides 
with the anti-Fokker-Planck equation, but otherwise 
it can not be used to calculate 
the negative Lyapunov exponents.

   The anti-Fokker-Planck equation is analogous to 
the "anti-Lorentz-Boltzmann equation" which was introduced to 
calculate the negative Lyapunov exponents 
using the kinetic approach \cite{Bei96,Bei98,Bei95}. 
   In this approach the anti-Lorentz-Boltzmann equation is given as 
the Lorentz-Boltzmann equation where the collision 
operator has the opposite sign to the ordinary 
Lorentz-Boltzmann equation 
in an equilibrium or a non-equilibrium stationary state. 

   At least, in the equilibrium case expressed by $\gamma=0$ and 
$\bar{\alpha}=0$ the anti-Fokker-Planck equation must provide the 
negative Lyapunov exponents as the opposites of the positive 
Lyapunov exponents calculated by using the Fokker-Planck 
equation (\ref{Fokke}). 
   In the next section we show it actually as the special case of 
more general results.


\section{Conjugate Pairing Rule 
for Thermostatted Systems with a Shear 
}

   We have to know the time evolution of the amplitude of the 
tangent vector in order to calculate the Lyapunov exponents. 
   Such a time evolution for the forward movement 
(the time-reversed movement)  is expressed as the dynamics of 
the diagonal elements of the matrix 
$\Upsilon^{(+)}(t)$ (the matrix $\Upsilon^{(-)}(t)$) given by 

\begin{eqnarray}
   \Upsilon^{(\pm)}(t) \equiv \left\langle \delta\bfq \delta\bfq^{T} 
   \right\rangle_{t}^{(\pm)}.
\label{MatriUpsil}\end{eqnarray} 

\noindent Here the bracket $\langle \cdots \rangle_{t}^{(\pm)}$ 
means to take the average by the probability density 
$\rho^{(\pm)}(\delta\bfGamma,t)$, namely 

\begin{eqnarray}
   \left\langle X(\delta\bfGamma) \right\rangle_{t}^{(\pm)} 
   \equiv \int d\delta\bfGamma X(\delta\bfGamma) \rho^{(\pm)}(\delta\bfGamma,t) 
\end{eqnarray} 

\noindent for any function $X(\delta\bfGamma)$ of 
$\delta\bfGamma$. 
   In Ref. \cite{Tan01} the positive Lyapunov exponents 
were calculated 
by the time-averaged exponential rate of the diagonal elements of 
an orthogonal-transformed matrix of $\Upsilon^{(+)}(t)$. 
   We get the negative Lyapunov exponents from the matrix 
$\Upsilon^{(-)}(t)$ by a similar procedure. 
   The introduction of Lyapunov exponents by the spatial 
coordinate part only (or the momentum part only) of 
the tangent vector has been used 
previously by Refs. \cite{Bei98,Zon98}.

   We introduce the matrix $\tilde{\Upsilon}^{(\pm)}(t)$ defined by 

\begin{eqnarray}
   \tilde{\Upsilon}^{(\pm)}(t)\equiv\Upsilon^{(\pm)}(\pm t) 
   e^{\pm\bar{\alpha}t } .
\label{MatriR}\end{eqnarray} 

\noindent  As shown in Appendix \ref{Lya}, 
the matrix $\tilde{\Upsilon}^{(\pm)}(t)$ satisfies 
the differential equation

\begin{eqnarray}
   &&\frac{d^4 \tilde{\Upsilon}^{(\pm)}(t)}{dt^4} 
      - \frac{1}{2} \left[ \Omega^2 
      \frac{d^2 \tilde{\Upsilon}^{(\pm)}(t)}{dt^2} 
      + \frac{d^2 \tilde{\Upsilon}^{(\pm)}(t)}{dt^2} 
      \left(\Omega^2 \right)^{T} \right] 
      \nonumber \\
   && \hspace{0.5cm} + \frac{1}{16}\left[ \Omega^4  
      \tilde{\Upsilon}^{(\pm)}(t)  
      -2 \Omega^2  \tilde{\Upsilon}^{(\pm)}(t) 
	  \left(\Omega^2\right)^{T} \right. 
      \nonumber \\
   && \hspace{1.5cm} \left.
      +\tilde{\Upsilon}^{(\pm)}(t) \left(\Omega^4 \right)^{T} \right] 
   - \frac{2}{m^{2}} \hat{{\cal D}} 
      \left\{ \frac{d \tilde{\Upsilon}^{(\pm)}(t)}{dt}\right\}
      \nonumber \\
   && \hspace{0.5cm}  = 0
\label{EquatR} \end{eqnarray} 

\noindent where we assumed the probability density 
$\rho^{(\pm)}(\delta\bfGamma,t)$ to be zero at the boundary of the 
tangent space.
   Here $\Omega$ is the $(2N) \times (2N)$ matrix defined by 

\begin{eqnarray}
   \Omega \equiv \Phi - \Psi . 
\label{MatriOmega}\end{eqnarray} 

\noindent and $\hat{{\cal D}}$ is the linear operator 
to map any $(2N) \times (2N)$ matrix 
$X\equiv(X_{jk})$ to the $(2N) \times (2N)$ matrix $\hat{{\cal D}} 
\{X \} 
\equiv (\; (\hat{{\cal D}} \{X \})_{jk}\;)$ defined by 

\begin{eqnarray}
   (\hat{{\cal D}} \{X \})_{jk} \equiv \sum_{\mu=1}^{N} 
   \sum_{\nu=1}^{N} 
   D_{j\mu k \nu} X_{\mu\nu} . 
\label{OperaD}\end{eqnarray} 

\noindent  It may be noted that Eq. (\ref{EquatR}) for the matrix 
$\tilde{\Upsilon}^{(\pm)}(t)$ is invariant under the transformation 
$\Omega\rightarrow-\Omega$.

   We can choose the initial probability density 
$\rho^{(+)}(\delta\bfGamma,0)$ arbitrarily 
to calculate the positive Lyapunov exponents. 
   On the other hand 
in order to derive the corresponding negative Lyapunov exponents
we assume the initial probability density 
$\rho^{(-)}(\delta\bfGamma,0)$ for the time-reversed tangent vector 
to satisfy the condition  
$d^{k} \tilde{\Upsilon}^{(-)}(t) / dt^{k} |_{t=0} 
= d^{k} \tilde{\Upsilon}^{(+)}(t) / dt^{k}|_{t=0}$, $k=0,1,2,3$ 
at the initial time $t=0$.
   Under this assumption it follows from Eq. (\ref{EquatR}) that

\begin{eqnarray}
   \Upsilon^{(-)}(- t)  e^{-\bar{\alpha}t} 
   = \Upsilon^{(+)}( t) e^{\bar{\alpha}t } , 
\label{RelatPosNeg1}\end{eqnarray} 

\noindent because the quantities $\tilde{\Upsilon}^{(+)}(t)$ 
and $\tilde{\Upsilon}^{(-)}(t)$ defined by Eq. (\ref{MatriR}) 
satisfy the same differential equation (\ref{EquatR}) 
and have the same initial condition.
   Therefore the diagonal element 
$\acute{\Upsilon}_{jj}^{(\pm)}(t)$ of 
{\em any} orthogonal-transformed matrix of $\Upsilon^{(\pm)}(t)$ 
must satisfy the relation 

\begin{eqnarray}
   \acute{\Upsilon}_{jj}^{(-)}(- t)  
   = \acute{\Upsilon}_{jj}^{(+)}( t) 
   e^{2\bar{\alpha}t } .
\label{RelatPosNeg2}\end{eqnarray} 

\noindent This equation connects the time-forward evolution and 
the time-reversed evolution in the amplitudes of the spatial part 
of the tangent vector in the thermostatted system. 

   The $j$-th positive (or zero) Lyapunov exponent $\lambda_{j}^{(+)}$ 
and its conjugate negative (or zero) exponent $\lambda_{j}^{(-)}$ are  
given by 

\begin{eqnarray}
   \lambda_{j}^{(\pm)} = \lim_{t=\rightarrow\pm\infty} 
   \frac{1}{2t} \ln 
   \frac{\acute{\Upsilon}_{jj}^{(\pm)}(t)}
   {\acute{\Upsilon}_{jj}^{(\pm)}(0)} .
\label{LyapuDefin}\end{eqnarray} 

\noindent The quantity $\acute{\Upsilon}_{jj}^{(\pm)}(t)$ 
satisfies the condition  
$\acute{\Upsilon}_{jj}^{(+)}(0)=\acute{\Upsilon}_{jj}^{(-)}(0)$ 
at the initial time, and  
using Eqs. (\ref{RelatPosNeg2}) and  (\ref{LyapuDefin}) 
we obtain 

\begin{eqnarray}
   \lambda_{j}^{(+)} + \lambda_{j}^{(-)} = - \bar{\alpha} . 
\label{PairiRule}\end{eqnarray} 

\noindent This is the conjugate pairing rule of the Lyapunov spectrum 
for the iso-kinetic thermostatted system with a shear field. 
   It is clear that it is attributed to the pairing rule of the  
Hamiltonian system in the case of $\bar{\alpha}=0$.


\section{Conjugate pairing rule for a color Field}

   For an actual calculation of the Lyapunov spectrum 
for an iso-kinetic thermostatted system by the master equation approach 
we have to know the value of tensor $D_{jkln}$ and to solve the 
differential equation (\ref{EquatR}) for the matrix 
$\tilde{\Upsilon}^{(\pm)}(t)$. 
   Let's discuss these points briefly by using a 
case without a shear field such as a color field system  
\cite{Del96,Pos88}, namely 

\begin{eqnarray}
   \gamma = 0. 
\label{GammaZero}\end{eqnarray}  

\noindent For simplicity in this section we also assume 
that the tensor $D_{jkln}$ 
is expressed as the multiplication of the matrix elements  of 
a symmetric $(2N) \times (2N)$ matrix $W\equiv (W_{jk})$: 

\begin{eqnarray}
   D_{jkln} = W_{jk} W_{ln}. 
\label{AssumMulti}\end{eqnarray} 

\noindent This assumption has already been used 
in Ref. \cite{Tan01}. 

   As shown in Appendix \ref{NoShear}, 
under the assumptions (\ref{GammaZero}) and (\ref{AssumMulti}) 
the equation 
of the matrix  $\tilde{\Upsilon}^{(\pm)}(t)$ is simplified to 

\begin{eqnarray}
   \frac{d^{3} \tilde{\Upsilon}^{(\pm)}(t)}{dt^{3}} 
   - \bar{\alpha}^{2} 
   \frac{d \tilde{\Upsilon}^{(\pm)}(t)}{dt} - 
   \frac{2}{m^{2}}  W \tilde{\Upsilon}^{(\pm)}(t) W =0. 
\label{EquatUpsilNoShear}\end{eqnarray} 

\noindent It is shown that Eq. (\ref{EquatR}) is  
given by taking the time-differential 
in both the sides of Eq. (\ref{EquatUpsilNoShear}). 
   Using the orthogonal matrix $V$ diagonalizing the matrix $W$, 
namely  

\begin{eqnarray}
   (V^{T} W V)_{jk} = \omega_{j}\delta_{jk} 
\end{eqnarray} 
  
\noindent with a real eigenvalue $\omega_{j}$, 
the quantity $\acute{\Upsilon}_{jj}^{(\pm)}(t)$ is expressed as 
the diagonal element of the matrix  
$\acute{\Upsilon}^{(\pm)}(t)\equiv 
(\acute{\Upsilon}_{jk}^{(\pm)}(t))$ 
defined by   

\begin{eqnarray}
   \acute{\Upsilon}^{(\pm)}(t) = V^{T}\Upsilon^{(\pm)}(t) V.
\label{NoShearAcuteUpsil}\end{eqnarray} 
 
%
  
\noindent Using Eq. (\ref{NoShearAcuteUpsil}) 
we can solve the equation for the quantity 
$\acute{\Upsilon}_{jj}^{(\pm)}(t)$ derived from 
Eq. (\ref{EquatUpsilNoShear}), and  by using  
Eqs. (\ref{MatriR}), (\ref{LyapuDefin}) 
and (\ref{NoShearAcuteUpsil}) 
the Lyapunov exponents are given by 

\begin{eqnarray}
   \lambda_{j}^{(\pm)} = -\frac{\bar{\alpha}}{2}
   \pm \frac{1}{2} \left(\Lambda_{j} 
   + \frac{\bar{\alpha}^{2}}{3\Lambda_{j}} \right) . 
\label{LyapuNoShear}\end{eqnarray} 

\noindent where $\Lambda_j$ is defined by 
   
\begin{eqnarray}
   \Lambda_{j} \equiv \left[ \;
    \left(\frac{\omega_{j}}{m}\right)^{2} + 
   \sqrt{\left(\frac{\omega_{j}}{m}\right)^{4} 
   -\left(\frac{\bar{\alpha}^{2}}{3} \right)^{3}} \; 
   \right]^{1/3} .
\label{NoShearLanbda}\end{eqnarray} 

\noindent 
(See Appendix \ref{NoShear} about a derivation of Eq. 
(\ref{LyapuNoShear}).)
   It is clear that the Lyapunov exponents given by Eq. 
(\ref{LyapuNoShear}) satisfy 
the conjugate pairing rule (\ref{PairiRule}). 

   Concerning the expression (\ref{LyapuNoShear}) 
for the Lyapunov exponent 
it is important to note that the tensor $D_{jkln}$ 
 can depend on 
external force fields.  
   This implies that the eigenvalue $\omega_{j}$ 
of the matrix $W$ can depend on the friction 
coefficient $\bar{\alpha}$.  
   If we were to assume the quantity $\omega_{j}$ 
to be independent of the friction coefficient 
$\bar{\alpha}$, then we obtain the expression of 
the Lyapunov exponents as 
$\lambda_{j}^{(\pm)} =\pm |\omega_{j}/(2m)|^{2/3} 
-\bar{\alpha}/2 +{\cal O}(\bar{\alpha}^{2})$ from 
Eq. (\ref{LyapuNoShear})  in the case of 
$|\bar{\alpha}| \leq \sqrt{3} \;|\omega_{j} / m |^{2/3}$.  
   However this is not consistent with the numerical 
results in a deterministic many-hard-disk system with 
a color field in which 
the negative Lyapunov exponents rather increase 
as the value of the friction coefficient increases 
\cite{Del96}.
   This consideration suggests that we should not neglect 
the external force field dependence of the correlation 
amplitude $D_{jkln}$ at least in the color field case.
   The dependence of the tensor $D_{jkln}$ on 
the shear rate and the external force fields 
should be a subject for a separated paper, 
although the conjugate pairing rule of the Lyapunov spectrum 
is correct regardless of their dependence as shown 
in this paper.


\section{Conclusion and Remarks}

   In this paper we have applied the master equation approach 
to Lyapunov spectra to non-equilibrium iso-kinetic thermostatted 
systems in order to discuss a conjugate pairing rule. 
   We considered two-dimensional many-particle system 
with Gaussian white random 
interactions between the particles. 
  In this system the positive Lyapunov 
exponents are calculated by a (forward) Fokker-Planck equation 
for the tangent vector dynamics. 
   We proposed a method to calculate the negative Lyapunov 
exponents by a time-reversed master equation, especially the 
anti-Fokker-Planck equation where the diffusion term has the 
opposite sign to the forward Fokker-Planck equation.  
   Using the Lyapunov exponents calculated by these two 
Fokker-Planck equations we showed the conjugate 
pairing rule of the Lyapunov spectrum for iso-kinetic thermostatted 
systems with a shear field given by the Sllod equation 
in the thermodynamic limit. 
   We also gave an concrete form to connect the Lyapunov 
exponents with the time-correlation of the interaction matrix 
in a thermostatted system without a shear field. 

   We discussed the conjugate pairing rule based on the iso-kinetic 
thermostat in the thermodynamic limit. 
   However it is known that the iso-kinetic thermostat is formally 
equivalent to other thermostats such as the iso-energetic thermostat 
in the thermodynamic limit \cite{Rue00}. 
   In this sense our result should be correct in systems 
with such other thermostats, more explicitly as far as 
the friction coefficient can be regarded as a constant 
even in a finite number of particle systems. 

   In order to get the ani-Fokker-Planck equation we used the fact 
that the shear rate $\gamma$ changes its sign in the time 
reversed motion. 
   However this time-reversed change of the sign of the shear rate 
to get the anti-Fokker-Planck equation may not be 
essential to obtain the negative Lyapunov exponents, 
if the Lyapunov exponents are even 
functions of the shear rate. 
   We have not proved the invariance of the Lyapunov exponents 
under the transformation $\gamma\rightarrow -\gamma$ in this paper, 
but the invariance of Eq. (\ref{EquatR}) under the transformation 
$\Omega\rightarrow -\Omega$ implies that the Lyapunov exponents 
are invariant under this transformation.  


   We can show that all the Lyapunov exponents $\lambda_{j}^{(+)}$ 
($\lambda_{j}^{(-)}$) are non-negative (non-positive) in the case 
of $\gamma=0$ (See Appendix \ref{NoShear}.). 
   This implies that in this case the number of the positive 
Lyapunov exponents should be equal to the number of the 
negative Lyapunov exponents. 
   However we have not proved that it is also correct in the 
presence of a shear field: $\gamma\neq0$. 
   Concerning this point we should notice that 
a numerical calculation of the Lyapunov spectrum for 
the Sllod equation (\ref{ThermEquat1}) and 
(\ref{ThermEquat2}) 
showed that in the case of a high shear rate the number of 
the positive Lyapunov exponents can be less than the 
number of the negative Lyapunov exponents \cite{Mor02}. 
   Therefore it should be interesting to check whether 
   the master equation approach to the Lyapunov spectrum 
can describe such a situation or not.  

   It should be noted that the discussion of this paper does not 
conclude that the conjugate pairing rule of the Lyapunov spectrum for 
the thermostatted system with a shear field must be satisfied rigorously 
in deterministic chaotic systems.    
   To show the conjugate pairing rule in this paper we 
assumed the Gaussian white randomness (\ref{WhiteNoise1}) 
and (\ref{WhiteNoise2}) for the particle interactions, and there is 
no guarantee that we can justify the conjugate pairing rule 
by the master 
equation approach under a more general random interaction of 
particles, especially under the randomness with a long time 
correlation which leads to a more general master equation 
for the tangent vector rather than a simple Fokker-Planck equation. 
   A generalization of the conjugate pairing rule by the 
master equation approach to a more general random 
interaction remains as one of the important future problems.


\section*{Acknowledgements}

We are grateful for financial support for this work 
from the Australian Research Council.


\appendix

 \setcounter{section}{0} 
 \makeatletter 
    \@addtoreset{equation}{section} 
    \makeatother 
    \def\theequation{\Alph{section}.%
    \arabic{equation}} 

\section{Master Equation for the Tangent Vector}
\label{Maste} 

In this appendix we derive the Fokker-Planck equation (\ref{Fokke}) 
for the tangent vector space. 
Using the Kramers-Moyal expansion the dynamics of the probability 
density $\rho^{(+)}(\delta\bfGamma,t)$ is given by \cite{Ris89} 

\begin{eqnarray}
   \frac{\partial \rho^{(+)}(\delta\bfGamma,t) }{\partial t} 
   &= & \sum_{n=1}^{\infty} \sum_{j_{1}=1}^{2N} \sum_{j_{2}=1}^{2N} 
   \cdots 
   \sum_{j_{n}=1}^{2N} (-1)^{n} \nonumber \\
   && \hspace{0.5cm} \times
   \frac{\partial^{n}\Xi_{j_{1}j_{2}\cdots j_{n}}^{(n)} 
   (\delta\bfGamma,t) 
   \rho^{(+)}(\delta\bfGamma,t)}{\partial\delta\caGamma_{j_{1}} 
   \partial\delta\caGamma_{j_{2}}\cdots\partial\delta\caGamma_{j_{n}}  
   }  
\label{Krame}\end{eqnarray}

\noindent where $\Xi_{j_{1}j_{2}\cdots 
j_{n}}^{(n)}(\delta\bfGamma,t)$ 
is defined by 

\begin{eqnarray}
    && \hspace{0cm} \Xi_{j_{1}j_{2} 
	   \cdots j_{n}}^{(n)}(\delta\bfGamma,t) \nonumber \\
   && \hspace{0.5cm}
   \equiv
      \frac{1}{n!} \lim_{s\rightarrow 0} \frac{1}{s} \Bigl\langle    
      [\delta\caGamma_{j_{1}}(t+s)-\delta\caGamma_{j_{1}}(t)]       
      \nonumber \\
   && \hspace{1.5cm}
      \times [\delta\caGamma_{j_{2}}(t+s)-\delta\caGamma_{j_{2}}(t)] 
      \nonumber \\
   && \hspace{1.5cm}  \left.\cdots   \times
      [\delta\caGamma_{j_{n}}(t+s)-\delta\caGamma_{j_{n}}(t)]
   \Bigr\rangle\right|_{\delta\smallbfGamma (t)=\delta\smallbfGamma}  
\label{ColliX}\end{eqnarray}

\noindent and $\delta\caGamma_{j}(t)$ is the $j$-th component of the 
tangent vector $\delta\bfGamma(t)$.

     Using  Eq. (\ref{LineaDynam}) we obtain

\begin{eqnarray}
    && \hspace{-0.5cm} \delta\bfGamma (t+s)-\delta\bfGamma (t) 
	\nonumber \\
   && \hspace{0cm}
    =\left\{\stackrel{\leftarrow}{T}  
   \exp{\left[\int_{t}^{t+s}d\tau\;{\cal L}(\tau)\right]}-1 
   \right\}\delta\bfGamma (t)
   \nonumber \\
   && \hspace{0cm}
    = \sum_{n=1}^{\infty} \int_{t}^{t+s}d\tau_{n}
    \int_{t}^{\tau_{n}}d\tau_{n-1}
    \cdots \int_{t}^{\tau_{2}}d\tau_{1} \; 
      \nonumber \\
   && \hspace{0cm}
     \hspace{1cm}\times
     {\cal L}(\tau_{n}){\cal L}(\tau_{n-1})\cdots {\cal L}(\tau_{1}) \delta\bfGamma (t) .
\label{TimeOrderExpa}\end{eqnarray}

\noindent It follows from Eqs. (\ref{MatriCalL}), (\ref{WhiteNoise1}), 
(\ref{WhiteNoise2}), (\ref{ColliX}) and (\ref{TimeOrderExpa}) that  

\begin{eqnarray}
    &&\bfXi^{(1)}(\delta\bfGamma,t) \nonumber \\
   &&\hspace{0.2cm} \equiv (\Xi_{1}^{(1)}(\delta\bfGamma,t), 
      \Xi_{2}^{(1)}(\delta\bfGamma,t), 
	  \cdots,\Xi_{2N}^{(1)}(\delta\bfGamma,t) )^{T}
      \nonumber \\
   &&\hspace{0.2cm} =  
      \lim_{s\rightarrow 0} \frac{1}{s}
	  \left. \bigl\langle \left[\delta\bfGamma(t+s)-\delta\bfGamma(t) 
	  \right] \bigr\rangle \right|_{\delta\smallbfGamma(t)
	  =\delta\smallbfGamma}
	  \nonumber \\
   &&\hspace{0.2cm} =
      \lim_{s\rightarrow 0} \frac{1}{s} \int_{t}^{t+s} d\tau 
	  \left\langle {\cal L}(\tau) \right\rangle \delta\bfGamma
	  \nonumber \\
   &&\hspace{0.2cm} = 
   \left(\begin{array}{c}
   \Phi\delta\bfq +\delta\bfp/m \\
      \Psi\delta\bfp/m 
   \end{array} \right)
\label{FunctX1}\end{eqnarray}

\noindent and

\begin{eqnarray}
   && \Xi^{(2)}(\delta\bfGamma,t) \equiv 
   (\;\Xi_{j k}^{(2)}(\delta\bfGamma,t) \;) \nonumber \\
   &&\hspace{0.2cm}= 
      \lim_{s\rightarrow 0} \frac{1}{2s}
	  \biggl\langle 
	     \left[\delta\bfGamma(t+s)-\delta\bfGamma(t) \right] 
	  \nonumber \\
   &&\hspace{2cm} \times 
	   \left. \left[\delta\bfGamma(t+s)-\delta\bfGamma(t) \right]^{T} 
	  \biggr\rangle \right|_{\delta\smallbfGamma(t)
	     =\delta\smallbfGamma}
	  \nonumber \\
   &&\hspace{0.2cm} =
      \lim_{s\rightarrow 0} \frac{1}{2s} 
	  \int_{t}^{t+s} d\kappa \int_{t}^{t+s} d\tau 
	  \nonumber \\
   &&\hspace{2cm} \times 
	  \left\langle {\cal L}(\kappa) \delta\bfGamma 
	  \delta\bfGamma^{T} {\cal L}(\tau)^{T} \right\rangle 
	  \nonumber \\
   &&\hspace{0.2cm} = \left(\begin{array}{cc} 
      0_{2N} & 0_{2N} \\
	  0_{2N} & \eta(\delta\bfq)
    \end{array} \right)
\label{FunctX2}\end{eqnarray}

\noindent where $\eta(\delta\bfq)\equiv 
(\eta_{jk}(\delta\bfq))$ is defined by 

\begin{eqnarray}
   \eta_{jk}(\delta\bfq)
   \equiv \frac{1}{2}
   \sum_{\mu=1}^{N} \sum_{\nu=1}^{N} 
	     D_{j\mu k\nu }\delta q_{\mu} \delta q_{\nu}.
\label{MatriEta}\end{eqnarray}

\noindent 
Here the only non-zero contributions come from the $n=1$ 
term of Eq. 
(\ref{TimeOrderExpa}). For general $n$, the number 
of delta functions
from Eq. (\ref{WhiteNoise2}) must be only one less than the number 
of time integrals, to give a non-zero contribution. 
   It is straightforward to show that this never happens for $n>1$.
   Concerning the terms including  $\Xi_{j_{1}j_{2}\cdots 
j_{n}}^{(n)}(\delta\bfGamma,t)$,  $n=3,4,\cdots$ in 
the right-hand side of Eq. (\ref{Krame}) we obtain 

\begin{eqnarray}
   \Xi_{j_{1}j_{2}\cdots 
   j_{n}}^{(n)}(\delta\bfGamma,t) = 0, 
   \;\;\;\;\;\;\;\; n=3,4,\cdots,
\label{HighterCoeff}\end{eqnarray}

\noindent because 
of the Gaussian white properties (\ref{WhiteNoise1}) and 
(\ref{WhiteNoise2}) of the random matrix $R(t)$. 
   Using Eqs.  (\ref{Krame}), 
(\ref{FunctX1}), (\ref{FunctX2}) and (\ref{HighterCoeff}) 
we obtain the Fokker-Planck equation (\ref{Fokke}). 


\section{Equation for the Matrix $\tilde{\Upsilon}^{(\pm)}$}
\label{Lya}

   In this appendix we give details of the derivation of 
Eq. (\ref{EquatR}) from Eqs. (\ref{Fokke}), 
(\ref{AntiFokke}) and (\ref{MatriUpsil}). 
   We start this derivation by introducing the $N \times N$ 
matrices ${\cal F}^{(\pm)}(t) $ and 
${\cal G}^{(\pm)}(t) $ defined by 
   
\begin{eqnarray}
   {\cal F}^{(\pm)}(t) \equiv \left\langle \delta\bfq \delta\bfp^{T} 
   \right\rangle_{t}^{(\pm)} ,
\label{MatriCalF}\end{eqnarray} 
\begin{eqnarray}
   {\cal G}^{(\pm)}(t) \equiv \left\langle \delta\bfp \delta\bfp^{T} 
   \right\rangle_{t}^{(\pm)} .
\label{MatriCalG}\end{eqnarray} 

\noindent    Equations (\ref{Fokke}) and (\ref{AntiFokke}) lead to 
the equations 

\begin{eqnarray}
   \frac{d \Upsilon^{(\pm)}(t)}{dt} &=& 
      \Phi \Upsilon^{(\pm)}(t) + \Upsilon^{(\pm)}(t) \Phi^{T} 
      \nonumber \\ 
   && + \frac{1}{m} 
      \left[ {\cal F}^{(\pm)}(t) + {\cal F}^{(\pm)}(t){}^{T}  \right] , 
\label{ConneUpsil} \end{eqnarray} 
\begin{eqnarray}
   \frac{d {\cal F}^{(\pm)}(t)}{dt} &=& 
      \Phi {\cal F}^{(\pm)}(t)  + {\cal F}^{(\pm)}(t)  \Psi^{T} 
 + \frac{1}{m} {\cal G}^{(\pm)}(t) , 
 \nonumber \\ 
   &&     
\label{ConneCalF}\end{eqnarray} 
\begin{eqnarray}
   \frac{d {\cal G}^{(\pm)}(t) }{dt} 
      &=& \Psi {\cal G}^{(\pm)}(t) + {\cal G}^{(\pm)}(t)  \Psi^{T} 
	   \pm 
      \hat{{\cal D}} \left\{\Upsilon^{(\pm)}(t) \right\}  \nonumber \\ 
   &&
\label{ConneCalG}\end{eqnarray}

\noindent for the matrices $\Upsilon^{(\pm)}(t)$, 
${\cal F}^{(\pm)}(t) $ 
and ${\cal G}^{(\pm)}(t) $ with the operator $\hat{{\cal D}}$ 
defined by 
Eq. (\ref{OperaD}). 
   Here, to derive Eq. (\ref{ConneCalG}) we used the relation 

\begin{eqnarray}
   D_{lnjk} = D_{jkln}
\end{eqnarray}

\noindent which is derived from 
the definition (\ref{WhiteNoise2}) of the tensor 
$D_{jkln}$ and the symmetry property of the matrix $R(t)$.   

   Eqs. (\ref{ConneUpsil}), (\ref{ConneCalF}) and (\ref{ConneCalG}) 
are equivalent to 

\begin{eqnarray}
   \frac{d \breve{\Upsilon}^{(\pm)}(t)}{dt} = 
      \breve{{\cal F}}^{(\pm)}(t) P(t) ^{T} 
	  + P(t)\breve{{\cal F}}^{(\pm) }(t){}^{T} ,
\label{ConneUpsil2}\end{eqnarray}
\begin{eqnarray}	  
   \frac{d \breve{{\cal F}}^{(\pm)}(t)}{dt} = 
      P(t) 
      \breve{{\cal G}}^{(\pm)}(t) , 
\label{ConneCalF2}\end{eqnarray}
\begin{eqnarray}	  
   \frac{d \breve{{\cal G}}^{(\pm)}(t) }{dt} = 
      \pm \hat{\tilde{{\cal D}}}_{t} 
	  \left\{\breve{\Upsilon}^{(\pm)}(t) \right\}
\label{ConneCalG2}\end{eqnarray}

\noindent where $\breve{\Upsilon}^{(\pm)}(t)$, 
$\breve{{\cal F}}^{(\pm)}(t)$ 
and $\breve{{\cal G}}^{(\pm)}(t)$ are defined by 
   
\begin{eqnarray}
   \breve{\Upsilon}^{(\pm)}(t)  &\equiv& 
      e^{-\Phi t} \Upsilon^{(\pm)}(t)  e^{-\Phi^{T} t} , 
\label{MatriBreveUpsil}\end{eqnarray}
\begin{eqnarray}
   \breve{{\cal F}}^{(\pm)}(t)  &\equiv& 
      e^{-\Phi t}{\cal F}^{(\pm)}(t)  e^{-\Psi^{T} t} , 
\label{MatriBreveF}\end{eqnarray}
\begin{eqnarray}
   \breve{{\cal G}}^{(\pm)}(t)  &\equiv& 
      e^{-\Psi t} {\cal G}^{(\pm)}(t)  e^{-\Psi^{T} t} , 
\label{MatriBreveG}\end{eqnarray}

\noindent and $P(t)$ is the $(2N) \times (2N)$ matrix defined by 

 \begin{eqnarray}
   P(t) \equiv \frac{1}{m} e^{-\Omega t}
\label{MatriP}\end{eqnarray}
  
\noindent with the matrix $\Omega$ defined by Eq. (\ref{MatriOmega}), 
and $\hat{\tilde{D}}_{t}\{\cdots\}$ is defined by 

\begin{eqnarray}
   \hat{\tilde{{\cal D}}}_{t} \{X \} \equiv 
   e^{-\Psi t} 
   \hat{{\cal D}} \left\{e^{\Phi t} X  e^{\Phi^{T} t}  \right\}
   e^{-\Psi^{T} t} 
\label{OperaTildeD}\end{eqnarray} 

\noindent for any $(2N) \times (2N)$ matrix $X$. 
   Here we used the relation 

\begin{eqnarray}
   \Phi\Psi=\Psi\Phi, 
\end{eqnarray} 

\noindent so that we have the equation $\exp\{-\Phi t\} \cdot 
\exp\{\Psi t\}=\exp\{-(\Phi -\Psi )t\}$.

   Noting that the matrix $\breve{{\cal G}}^{(\pm)}(t)$ is 
symmetric and the inverse matrix of the matrix $P(t)$ is given 
by $P(t)^{-1}=m \exp\{\Omega t\}$, 
we obtain  

\begin{eqnarray}
   2 \breve{{\cal G}}^{(\pm)}(t) 
   &=& P(t)^{-1}  \frac{d \breve{{\cal F}}^{(\pm)}(t)}{dt} 
      +  \frac{d \breve{{\cal F}}^{(\pm)}(t){}^{T}}{dt}  
	   \left[P(t)^{-1}\right]^{T}  \nonumber \\
   &=&  P(t)^{-1}  \frac{d^{2} \breve{\Upsilon}^{(\pm)}(t)}{dt^{2}} 
	   \left[P(t)^{-1}\right]^{T} 
	   \nonumber \\
   && \hspace{1cm} + 
      P(t)^{-1}  \breve{{\cal F}}^{(\pm)}(t) \Omega^{T} 
	   \nonumber \\
   && \hspace{1cm} 
      +\Omega\breve{{\cal F}}^{(\pm)}(t){}^{T}
      \left[P(t)^{-1}\right]^{T} 
\label{ConneCalFUpsil} \end{eqnarray} 
      
\noindent by using Eqs. (\ref{ConneUpsil2}),  (\ref{ConneCalF2}) and 
(\ref{MatriP}).  
   Besides, using Eqs. (\ref{ConneUpsil2}),  (\ref{ConneCalF2}) and 
(\ref{MatriP}) we obtain   
   
\begin{eqnarray}
   && \hspace{-1cm} \frac{d}{dt} \left\{P(t)^{-1}  
   \breve{{\cal F}}^{(\pm)}(t) \Omega^{T} 
      +\Omega\breve{{\cal F}}^{(\pm)}(t){}^{T}
      \left[P(t)^{-1}\right]^{T} \right\}  \nonumber \\
   && \hspace{0cm}=
	   \Omega P(t)^{-1} 
	   \frac{d\breve{\Upsilon}^{(\pm)}(t)}{dt} \left[P(t)^{-1}\right]^{T}
	   \Omega^{T}     \nonumber \\
   && \hspace{1cm} + \breve{{\cal G}}^{(\pm)}(t) \Omega^{T} 
      +\Omega\breve{{\cal G}}^{(\pm)}(t) 
\label{ConneCalFG} \end{eqnarray} 

\noindent where we again used the relation 
$\breve{{\cal G}}^{(\pm)}(t)^{T} =\breve{{\cal G}}^{(\pm)}(t) $.
   It follows from Eqs. (\ref{ConneUpsil2}), (\ref{ConneCalF2}) 
and (\ref{ConneCalG2}) that 

\begin{eqnarray}
   && \pm 2 \hat{\tilde{{\cal D}}}_{t} 
      \left\{\breve{\Upsilon}^{(\pm)}(t) \right\}
      \nonumber \\
   && \hspace{0.2cm}= \frac{d}{dt} 
       P(t)^{-1}  \frac{d^{2} \breve{\Upsilon}^{(\pm)}(t)}{dt^{2}} 
	   \left[P(t)^{-1}\right]^{T} 
	   \nonumber \\
   && \hspace{1cm} 
	   + \Omega P(t)^{-1} 
	   \frac{d\breve{\Upsilon}^{(\pm)}(t)}{dt} 
	   \left[P(t)^{-1}\right]^{T}
	   \Omega^{T} 	     \nonumber \\
   && \hspace{1cm} +  \breve{{\cal G}}^{(\pm)}(t) \Omega^{T} 
      +\Omega\breve{{\cal G}}^{(\pm)}(t) .
\label{EquatUpsilCalG}\end{eqnarray} 

\noindent Taking the time-differential of both the sides of 
Eq. (\ref{EquatUpsilCalG}), and using Eq. (\ref{ConneCalG2})
we obtain 

\begin{eqnarray}
   && \frac{d^{2}}{dt^{2}} 
       P(t)^{-1}  \frac{d^{2} \breve{\Upsilon}^{(\pm)}(t)}{dt^{2}} 
	   \left[P(t)^{-1}\right]^{T} 
	   \nonumber \\
   && \hspace{1cm} 
	   + \Omega\frac{d}{dt}  P(t)^{-1} 
	   \frac{d\breve{\Upsilon}^{(\pm)}(t)}{dt} 
	   \left[P(t)^{-1}\right]^{T}
	   \Omega^{T} \nonumber \\
   && \hspace{1cm} 
	  \mp 2 \frac{d}{dt} 
      \hat{\tilde{{\cal D}}}_{t} 
	  \left\{\breve{\Upsilon}^{(\pm)}(t) \right\}
	  \nonumber \\
   && \hspace{1cm} \pm \hat{\tilde{{\cal D}}}_{t} 
      \left\{\breve{\Upsilon}^{(\pm)}(t) \right\}\Omega^{T} 
      \pm \Omega\hat{\tilde{{\cal D}}}_{t} 
	  \left\{\breve{\Upsilon}^{(\pm)}(t) \right\}
	     \nonumber \\
   && \hspace{1cm} =0 .
\label{EquatUpsil}\end{eqnarray} 

\noindent This is the equation 
for the quantity $\breve{\Upsilon}^{(\pm)}(t)$ only. 
	  
   Now we derive  the equation for  
$\tilde{\Upsilon}^{(\pm)}(t)$ defined 
by Eq. (\ref{MatriR}) from Eq. (\ref{EquatUpsil}) for 
$\breve{\Upsilon}^{(\pm)}(t)$ defined by 
Eq. (\ref{MatriBreveUpsil}). 
   We note 
   
\begin{eqnarray}
   \Omega &=& 2\Phi + \bar{\alpha} I_{2N}
   \label{ConneOmega} \\
   &=& -2\Psi - \bar{\alpha} I_{2N}
   \label{ConneOmega2}
\end{eqnarray}

\noindent which is derived from Eqs. (\ref{MatriPhi}), 
(\ref{MatriPsi}) and (\ref{MatriOmega}).    
   Using Eqs. (\ref{MatriR}), (\ref{MatriBreveUpsil}) 
and (\ref{ConneOmega}) 
the matrices $\breve{\Upsilon}^{(\pm)}(t)$ are 
connected with the matrix $\tilde{\Upsilon}^{(\pm)}(t)$ by
   
\begin{eqnarray}
   \breve{\Upsilon}^{(\pm)}(t) 
   = e^{-\Omega t/2} \tilde{\Upsilon}^{(\pm)}(\pm t) 
   e^{-\Omega^{T} t/2} .  
\label{ConneUpsilR}\end{eqnarray}

\noindent We introduce the multiplication 
$X\otimes Y$ of $X$ and $Y$ which is defined by

\begin{eqnarray}  
   X\otimes Y \equiv \frac{1}{2} \left[XY+(XY)^{T} \right]
\label{Multi}\end{eqnarray}

\noindent for any square matrices $X$ and $Y$ of the same size.
   This multiplication is used in the relation 
   
\begin{eqnarray}
   && \frac{d}{dt}e^{\pm\Omega t/2} Z(t) e^{\pm\Omega^{T} t/2} 
   \nonumber \\
   && = e^{\pm\Omega t/2} \left[ 
   \frac{d Z(t)}{dt} \pm \Omega\otimes Z(t) 
   \right] e^{\pm\Omega^{T} t/2} 
\label{DerivOtime}\end{eqnarray}  
   
\noindent satisfied by any $(2N)\times(2N)$ 
symmetric matrix $Z(t)$ 
as a function of $t$.   
   Noting Eqs. (\ref{Multi}) and (\ref{DerivOtime}), 
Eq. (\ref{ConneUpsilR}) leads to

\begin{eqnarray}  
   && P(t)^{-1} \frac{d \breve{\Upsilon}^{(\pm)}(t)}{dt} 
      \left[P(t)^{-1}\right]^{T} \nonumber \\
   && \hspace{0cm} = m^{2}e^{\Omega t/2} \left[ 
       \frac{d \tilde{\Upsilon}^{(\pm)}(\pm t)}{dt}
	  -\Omega\otimes\tilde{\Upsilon}^{(\pm)}(\pm t)
      \right] e^{\Omega^{T} t/2} ,
\label{FirstDerivU}\end{eqnarray}
\begin{eqnarray}  
   && P(t)^{-1} \frac{d^{2}\breve{\Upsilon}^{(\pm)}(t)}{dt^{2}} 
      \left[P(t)^{-1}\right]^{T} \nonumber \\
   && \hspace{0cm}= m^{2}e^{\Omega t/2} \left\{ 
      \frac{d^{2} \tilde{\Upsilon}^{(\pm)}(\pm t)}{dt^{2}}
	  -2\Omega\otimes\frac{d \tilde{\Upsilon}^{(\pm)}(\pm t)}{dt} 
	  \right. \nonumber \\
   && \hspace{2.5cm} + \Omega\otimes\left. \left[
	  \Omega\otimes\tilde{\Upsilon}^{(\pm)}(\pm t)\right]
      \right\} e^{\Omega^{T} t/2}  
\label{SeconDerivU}\end{eqnarray}

\noindent where we used the relation 
$\breve{\Upsilon}^{(\pm)}(t)^{T} = \breve{\Upsilon}^{(\pm)}(t)$. 
   Moreover $\hat{\tilde{{\cal D}}}_{t}$ 
operated on the matrix $\breve{\Upsilon}^{(\pm)}(t)$ is 
connected with $\hat{{\cal D}} $ 
operated on the matrix $\tilde{\Upsilon}^{(\pm)}(\pm t)$ as 

\begin{eqnarray}
   \hat{\tilde{{\cal D}}}_{t} 
   \left\{\breve{\Upsilon}^{(\pm)}(t) \right\}
   =  e^{\Omega t/2}  \hat{{\cal D}} 
   \left\{ \tilde{\Upsilon}^{(\pm)}(\pm t) \right\} 
   e^{\Omega^{T} t/2}  
\label{ConneTildeDD}\end{eqnarray}  

\noindent where we used Eqs.  (\ref{MatriR}), (\ref{MatriBreveUpsil})  
(\ref{OperaTildeD}) and (\ref{ConneOmega2}). 
   Inserting Eqs. (\ref{FirstDerivU}), 
(\ref{SeconDerivU}) and (\ref{ConneTildeDD}) 
into Eq. (\ref{EquatUpsil}) and using Eq. (\ref{DerivOtime}) 
we obtain 

\begin{eqnarray}
   && \frac{d^4 \tilde{\Upsilon}^{(\pm)}(\pm t)}{dt^4} 
      - 2 \Omega\otimes\left[\Omega\otimes
	  \frac{d^2 \tilde{\Upsilon}^{(\pm)}(\pm t)}{dt^2} \right]
	  \nonumber \\ 
   && \hspace{0.5cm} 
	  +\Omega\frac{d^2 \tilde{\Upsilon}^{(\pm)}(\pm t)}{dt^2}
	  \Omega^{T}
	  \nonumber \\ 
   && \hspace{0.5cm} 
	  +\Omega\otimes\left[\Omega\otimes\left[
	 \Omega\otimes\left[\Omega\otimes\tilde{\Upsilon}^{(\pm)}(\pm t)
	 \right]\right]\right]
	  \nonumber \\ 
   && \hspace{0.5cm} 
	 -\Omega\left[\Omega\otimes\left[\Omega\otimes
	 \tilde{\Upsilon}^{(\pm)}(\pm t)\right]\right]
	 \Omega^{T}
	  \nonumber \\ 
   && \hspace{0.5cm} 
	  \mp \frac{2}{m^{2}} \hat{{\cal D}} 
      \left\{ \frac{d \tilde{\Upsilon}^{(\pm)}(\pm t)}{dt}\right\}
      \;\; = \;\; 0.
\label{EquatR0a}\end{eqnarray}  

\noindent Eq. (\ref{EquatR0a}) is equivalent to 

\begin{eqnarray}
   && \frac{d^4 \tilde{\Upsilon}^{(\pm)}(\pm t)}{dt^4} 
       \nonumber \\
   && \hspace{0.5cm}
      -\frac{1}{2}\left[ \Omega^2 
      \frac{d^2 \tilde{\Upsilon}^{(\pm)}(\pm t)}{dt^2} 
      +\frac{d^2 \tilde{\Upsilon}^{(\pm)}(\pm t)}{dt^2} 
      \left(\Omega^2 \right)^{T} \right] \nonumber \\
   && \hspace{0.5cm} + \frac{1}{16}\left[ \Omega^4  
      \tilde{\Upsilon}^{(\pm)}(\pm t)  
      -2 \Omega^2  \tilde{\Upsilon}^{(\pm)}(\pm t) 
	  \left(\Omega^2\right)^{T} 
	  \right. \nonumber \\ 
   && \hspace{1cm} \left. 
      +\tilde{\Upsilon}^{(\pm)}(\pm t) 
	  \left(\Omega^4 \right)^{T} \right] 
      \mp \frac{2}{m^{2}} \hat{{\cal D}} 
      \left\{ \frac{d \tilde{\Upsilon}^{(\pm)}(\pm t)}{dt}\right\}
      \nonumber \\
   && \hspace{0.5cm} = 0.
\label{EquatR0}\end{eqnarray}  

\noindent By exchanging $t$ with $\pm t$ in Eq. (\ref{EquatR0}) 
we obtain Eq. (\ref{EquatR}).


\section{Lyapunov Exponents in the color Field Case}
\label{NoShear}

In this appendix we consider the case with no shear 
field using the condition (\ref{GammaZero}), and derive 
Eq. (\ref{EquatUpsilNoShear}) under the assumption 
(\ref{AssumMulti}). 
   We also give a derivation of Eq. (\ref{LyapuNoShear}) 
briefly. 

   Under the condition (\ref{GammaZero}), 
the matrix $\Omega$ defined by Eq. (\ref{MatriOmega}) 
is simply an identical matrix multiplied 
by a constant and is given by 

\begin{eqnarray}
   \Omega = \bar{\alpha} I_{2N}, 
\end{eqnarray}

\noindent and the matrix $P(t)$ defined 
by Eq. (\ref{MatriP}) and the operator 
$\hat{\tilde{{\cal D}}}_{t}\{\cdots\}$ defined by 
Eq. (\ref{OperaTildeD}) are given by 

\begin{eqnarray}
   P(t) = \frac{1}{m} e^{-\bar{\alpha}t} I_{2N}, 
\label{NoshearP}\end{eqnarray}
\begin{eqnarray}
   \hat{\tilde{{\cal D}}}_{t} \left\{ X \right\}
   = e^{2\bar{\alpha}t} 
   \hat{{\cal D}} \left\{ X \right\}
\label{NoshearD}\end{eqnarray}

\noindent for any $(2N)\times(2N)$ matrix $X$. 
   Noting Eqs. (\ref{NoshearP}) and 
(\ref{NoshearD}) and the relation 
$\breve{\Upsilon}^{(\pm)}(t)=\Upsilon^{(\pm)}(t)$, 
Eqs. (\ref{ConneUpsil2}), (\ref{ConneCalF2}) 
and (\ref{ConneCalG2}) are simply attributed into 

\begin{eqnarray}
   \frac{d \Upsilon^{(\pm)}(t)}{dt} 
   = \frac{1}{m} e^{-\bar{\alpha}t}
      \left[\breve{{\cal F}}^{(\pm)}(t) 
	  + \breve{{\cal F}}^{(\pm) }(t){}^{T}
	  \right] , 
	  \label{ConneUpsil3}\end{eqnarray}  
\begin{eqnarray}
   \frac{d \breve{{\cal F}}^{(\pm)}(t)}{dt} = 
      \frac{1}{m}
      e^{-\bar{\alpha}t} \breve{{\cal G}}^{(\pm)}(t) , 
	  \label{ConneCalF3}\end{eqnarray}  
\begin{eqnarray}	  
   \frac{d \breve{{\cal G}}^{(\pm)}(t) }{dt} = 
      \pm e^{2\bar{\alpha}t}
      \hat{{\cal D}} \left\{\Upsilon^{(\pm)}(t) \right\}.
	  \label{ConneCalG3}
\end{eqnarray}

\noindent Noting that the matrix $\breve{{\cal G}}^{(\pm)}(t)$ 
is symmetric, Eqs. (\ref{ConneUpsil3}), (\ref{ConneCalF3}) 
and (\ref{ConneCalG3}) lead to the differential equation

\begin{eqnarray}
  \frac{d}{dt} e^{\bar{\alpha}t}
  \frac{d}{dt} e^{\bar{\alpha}t}
  \frac{d}{dt} \Upsilon^{(\pm)}(t) = 
      \pm \frac{2}{m^{2}} 
  e^{2\bar{\alpha}t} \hat{{\cal D}}
   \left\{\Upsilon^{(\pm)}(t) \right\}
\label{EquatUpsilNoShear2}\end{eqnarray}  
	  
\noindent for the matrix $\Upsilon^{(\pm)}(t)$ only. 

   Now we consider the derivation of the equation for the matrix 
$\tilde{\Upsilon}^{(\pm)}(t)$ (defined by 
Eq. (\ref{MatriR})) from Eq. (\ref{EquatUpsilNoShear2}) 
for the matrix $\Upsilon^{(\pm)}(t)$.  
   It follows from Eqs. (\ref{MatriR}) 
and (\ref{EquatUpsilNoShear2}) that 

\begin{eqnarray}
   && \frac{d^{3} \tilde{\Upsilon}^{(\pm)}(\pm t)}{dt^{3}}
   - \bar{\alpha}^{2} 
   \frac{d \tilde{\Upsilon}^{(\pm)}(\pm t)}{dt} 
   \mp \frac{2}{m^{2}} \hat{{\cal D}} \left\{ 
   \tilde{\Upsilon}^{(\pm)}(\pm t) \right\} = 0 . 
    \nonumber \\ 
   &&     
\label{EquatUpsilNoShear3}\end{eqnarray} 

\noindent By exchanging $t$ with $\pm t$ 
in Eq. (\ref{EquatUpsilNoShear3}) 
and using Eqs. (\ref{OperaD}) and (\ref{AssumMulti}) we obtain 
Eq. (\ref{EquatUpsilNoShear}).


   Using Eqs. (\ref{EquatUpsilNoShear}) 
and (\ref{NoShearAcuteUpsil}) the real   
function $
\xi_{j}^{(\pm)}(t)$ of $t$ defined by 

\begin{eqnarray}
\xi_{j}^{(\pm)}(t)\equiv (V^{T} 
\tilde{\Upsilon}_{jj}^{(\pm)}(t) V )_{jj} =
\acute{\Upsilon}_{jj}^{(\pm)}(\pm t)e^{\pm\bar{\alpha}t}
\label{Xi}\end{eqnarray} 

\noindent satisfies 
the equation  

\begin{eqnarray}
   \frac{d^{3} \xi_{j}^{(\pm)}(t)}{dt^{3}} 
   - \bar{\alpha}^{2} 
   \frac{d \xi_{j}^{(\pm)}(t)}{dt} 
   - \frac{2\omega_{j}^{2} }{m^{2}}  
   \xi_{j}^{(\pm)}(t) =0. 
\label{EquatUpsilNoShear4}\end{eqnarray} 

\noindent The real solution of the linear differential 
equation (\ref{EquatUpsilNoShear4}) 
is expressed as 

\begin{eqnarray}
   \xi_{j}^{(\pm)}(t) = 
   \sum_{k=1}^{3} \mbox{Re} \left\{ 
   \upsilon_{j}^{(k)} e^{\zeta_{j}^{(k)} t} 
   \right\}
\label{EquatUpsilNoShearSolut}\end{eqnarray} 

\noindent where $\upsilon_{j}^{(k)}$, $j=1,2,3$ are constants 
determined by the initial condition, and $\zeta_{j}^{(k)}$, $
j=1,2,3$ are the three solutions of the equation 

\begin{eqnarray}
   && \zeta^{3} 
   - \bar{\alpha}^{2} \zeta   - 
   \frac{2\omega_{j}^{2} }{m^{2}}  = 0 
\label{3rdEquat}\end{eqnarray} 

\noindent for $\zeta$. 
   Here $\mbox{Re}\{X\}$ means to take the real part 
of any imaginary number $X$. 
   We sort the quantities $\zeta_{j}$, $
j=1,2,3$ as $\mbox{Re}\{\zeta_{j}^{(1)}\}\geq 
\mbox{Re}\{\zeta_{j}^{(2)}\}\geq \mbox{Re}\{\zeta_{j}^{(3)}\}$, 
so that using Eqs. (\ref{LyapuDefin}), (\ref{NoShearAcuteUpsil}) 
and (\ref{Xi})  
the Lyapunov exponent 
$\lambda_{j}^{(\pm)}$ is expressed as 

\begin{eqnarray}
   \lambda_{j}^{(\pm)} 
   &=& \pm \lim_{t=\rightarrow+\infty} 
   \frac{1}{2t} \ln 
   \frac{\xi_{j}^{(\pm)}(t)e^{\mp\bar{\alpha}t}}
   {\xi_{j}^{(\pm)}(0)} \nonumber \\
   &=& - \frac{\bar{\alpha}}{2} 
   \pm \frac{1}{2} \mbox{Re}\{\zeta_{j}^{(1)}\}. 
\label{NoShearLyapu}\end{eqnarray} 

\noindent It follows from Eq. (\ref{3rdEquat}) that 
$\zeta=\zeta_{j}^{(1)}$ 
is a real solution of Eq.  (\ref{3rdEquat}) and satisfies  
the conditions $\zeta_{j}^{(1)}\geq|\bar{\alpha}|$ 
$\lim_{\omega_{j}\rightarrow 0}\zeta_{j}^{(1)}=|\bar{\alpha}|$, 
noting that the point $\zeta = \zeta_{j}^{(1)}$ 
is the maximum intersecting point of the graphs  
$y = \zeta^{3} - \bar{\alpha}^{2} \zeta$ and 
$y = 2\omega_{j}^{2}/m^{2}$ in the $\zeta$-$y$ plain. 
   This means that the Lyapunov exponents 
$\lambda_{j}^{(+)}$ ($\lambda_{j}^{(-)}$) 
must be non-negative (non-positive). 
   More concretely the quantity 
$\zeta_{j}^{(1)}$ is given by 
 
\begin{eqnarray}
   \zeta_{j}^{(1)}
   = \Lambda_{j} + \frac{\bar{\alpha}^{2}}{3\Lambda_{j}}
\label{LyapuNoShear2}\end{eqnarray} 

\noindent with the quantity $\Lambda_{j}$ defined by Eq. 
(\ref{NoShearLanbda}).  
   We can check that in the case of $|\bar{\alpha}|\leq 
\sqrt{3}\;|\omega_{j}/m|^{2/3}$ the 
quantities $\Lambda_{j}$ and $\zeta_{j}^{(1)}$ 
are both real numbers, and in the case of $|\bar{\alpha}| >  
\sqrt{3}\;|\omega_{j}/m|^{2/3}$ the quantity $\Lambda_{j}$ 
can be an imaginary number but the quantity $\zeta_{j}^{(1)}$ 
given by Eq. (\ref{LyapuNoShear2}) 
is still a real number and satisfies the condition 
$\lim_{\omega_{j}\rightarrow 0}\zeta_{j}^{(1)}=|\bar{\alpha}|$. 
   Using Eqs. (\ref{NoShearLyapu}) and (\ref{LyapuNoShear2}) 
we obtain Eq. (\ref{LyapuNoShear}).



\end{multicols}

\end{document}